# The relationship between driving volatility in time to collision and crash-injury severity in a naturalistic driving environment

Behram Wali (Corresponding Author)
Graduate Research Assistant
Department of Civil and Environmental Engineering
The University of Tennessee, Knoxville
bwali@vols.utk.edu

Asad J. Khattak, Ph.D.
Beaman Distinguished Professor and Transportation Program Coordinator
Department of Civil and Environmental Engineering
The University of Tennessee, Knoxville
akhattak@utk.edu

Thomas Karnowski, Ph.D.
Research Staff Member
Imaging, Signals, and Machine Learning Group
Oak Ridge National Laboratory, TN, USA
karnowskitp@ornl.gov

# The relationship between driving volatility in time to collision and crash-injury severity in a naturalistic driving environment

**Abstract –**

As a key indicator of unsafe driving, driving volatility characterizes the variations in microscopic driving decisions. This study characterizes volatility in longitudinal and lateral driving decisions and examines the links between driving volatility in time to collision and crash-injury severity. By using a unique real-world naturalistic driving database from the 2nd Strategic Highway Research Program (SHRP), a test set of 671 crash events featuring around 0.2 million temporal samples of real-world driving are analyzed. Based on different driving performance measures, 16 different volatility indices are created. To explore the relationships between crash-injury severity outcomes and driving volatility, the volatility indices are then linked with individual crash events including information on crash severity, drivers' pre-crash maneuvers and behaviors, secondary tasks and durations, and other factors. As driving volatility prior to crash involvement can have different components, an in-depth analysis is conducted using the aggregate as well as segmented (based on time to collision) real-world driving data. To account for the issues of observed and unobserved heterogeneity, fixed and random parameter logit models with heterogeneity in parameter means and variances are estimated. The empirical results offer important insights regarding how driving volatility in time to collision relates to crash severity outcomes. Overall, statistically significant positive correlations are found between the aggregate (as well as segmented) volatility measures and crash severity outcomes. The findings suggest that greater driving volatility (both in longitudinal and lateral direction) in time to collision increases the likelihood of police reportable or most severe crash events. Importantly, compared to the effect of volatility in longitudinal acceleration on crash outcomes, the effect of volatility in longitudinal deceleration is significantly greater in magnitude. Methodologically, the random parameter models with heterogeneity-in-means and variances significantly outperformed both the fixed parameter and random parameter counterparts (with homogeneous means and variances), underscoring the importance of accounting for both observed and unobserved heterogeneity. The relevance of the findings to the development of proactive behavioral countermeasures for drivers is discussed.

# INTRODUCTION and BACKGROUND

The 2018 Traffic Safety Facts published by the National Highway Traffic Safety Administration (NHTSA) reported a total of 36,560 traffic fatalities and an additional 2,491,000 injuries in the U.S. (NHTSA, 2018). Of all the fatalities and injuries sustained by vehicle occupants, drivers sustained 75% of fatalities and 72% of injuries (NHTSA, 2018). As a result of extensive research over the decades (Mannering and Bhat, 2014), a broad spectrum of factors are known to be associated with the injury severity outcomes of drivers including drivers' characteristics, crash and roadway factors, vehicle features, weather, and environment-related factors (Kockelman and Kweon, 2002; Quddus et al., 2002; Abdel-Aty, 2003; Zajac and Ivan, 2003; Khattak and Targa, 2004; Mooradian et al., 2013; Behnood and Mannering, 2015; Ahmad et al., 2019b). Driving behavior, however, is a leading contributor to the occurrence of crashes and the injuries/fatalities therein. A better understanding of driving behavior prior to involvement in a crash is fundamental to the design of behavioral countermeasures. Thus, an analysis of the behavioral correlates of crash-injury severity has been of interest (Abdel-Aty, 2003; Paleti et al., 2010; Zhu and Srinivasan, 2011). Primarily, the focus has been on what is referred to as "aggressive" driving (such as driver was speeding, tailgating, improper lane changes, making obscene gestures, and so on), and its correlation with injury outcomes[1] (Nevarez et al., 2009; Richards and Cuerden, 2009; Paleti et al., 2010; Weiss et al., 2014). By using "aggressive" driving as a latent construct, Paleti et al. (2010) quantified the moderating effect of aggressive driving in increasing injury severity outcomes (Paleti et al., 2010). Likewise, as surrogates of driving behavior, higher speeds or speed limits are known to be correlated with higher injury severity outcomes (Duncan et al., 1998; Klop and Khattak, 1999; Renski et al., 1999; Abdel-Aty, 2003; Quddus et al., 2009; Weiss et al., 2014).

In the majority of the literature, crash causation studies or police crash reports have been used to gain an understanding of the relationships between crash-injury outcomes and driver-specific behavioral factors (Paleti et al., 2010; Savolainen et al., 2011; Mannering and Bhat, 2014). As acknowledged in the literature (Paleti et al., 2010; Mannering and Bhat, 2014; Imprialou and Quddus, 2019), classifying "aggressive" driving based on information (such as speeds, maneuvers, etc.) in police crash reports is a subjective process and there exists the possibility of misclassification. Also, the extent to which the speed information in police crash reports, typically used as a measure of driving behavior, is accurate is unclear. Importantly, while analysis of such a nature has helped to formulate actionable strategies for development of behavioral countermeasures, it does not capture the real-world instantaneous decisions or actual driving tasks immediately prior to a crash (Kim et al., 2016). Having said this, a deeper analysis is required to gain an understanding of how real-world microscopic driving decisions (e.g., speed, accelerations, vehicular jerk, etc.) in time to collision correlate with crash-injury outcomes. With the advent of naturalistic driving data, such an analysis is now possible.

---

[1] There exists extensive psychometrics-related literature regarding latent constructs for characterizing aggressive driving (Shinar, 1998). However, for brevity we focus on studies linking driving behavior explicitly with injury outcomes.

## Concept of Driving Volatility

Thanks to maniaturization and the rapid advancements in ambient sensing related technologies, the combination of sensing, computation, and communication now allows collection of great amounts of spatiotemporal behavioral data in an unintrusive manner. With the integration of communication technologies with video and radar surveillance, enormous magnitudes of real-world contextual driving data are now easily available (Campbell, 2012; Henclewood, 2014). The real-world large-scale driving data generated by advanced technologies are not informative to drivers in the raw form (Khattak and Wali, 2017). However, the raw information can be transformed into useful and actionable knowledge with appropriate data mining techniques – allowing a deeper and richer understanding of instantaneous driving decisions (Khattak and Wali, 2017). Important in this regard is the concept of "driving volatility" that captures the extent of variations in driving, especially hard accelerations/braking and jerky maneuvers (Liu et al., 2015; Wang et al., 2015b; Liu and Khattak, 2016; Liu et al., 2017). As a key measure of driving performance – it characterizes extreme behaviors and variations in real-world driving decisions (Liu and Khattak, 2016; Khattak and Wali, 2017). Compared to traditional surrogates of driving behavior (such as speed and driver demographics), the idea of driving volatility allows the development of proactive and more personalized warning and control assist systems[2] (Liu and Khattak, 2016; Khattak and Wali, 2017).

## Driving Volatility and Safety

An in-depth examination of short-term driving decisions in time to collision can shed light on the actual mechanism in which a vehicle is maneuvered or operated before a crash. While previous research used rigorous data mining techniques to characterize volatility in driving behaviors (Liu et al., 2015; Wang et al., 2015b; Liu and Khattak, 2016; Khattak and Wali, 2017; Liu et al., 2017), the real-world driving volatility metrics were not linked with unsafe outcomes. To this end, the concept of "driving volatility matrix" enables conceptualization of the extent of variations in driving behaviors at multiple levels of the transportation ecosystem and integrates real-world driving volatility with safety-critical events (e.g., crashes/near-crashes) (Wali et al., 2019b). Along these lines, previous studies extended driving volatility concept to specific locations and proposed a methodology for linking high frequency microscopic connected vehicles driving data with police-reported crashes (Kamrani et al., 2017; Wali et al., 2018b). Likewise, Kim et al. (2016) analyzed the links between micro-scale driving behavior and rear-end crash propensity in an exploratory manner (Kim et al., 2016). Extreme rates of deceleration were correlated with crashes occurring at signalized intersections (Wali et al., 2018b) and freeway ramp related rear-end crashes (Kim et al., 2016). Using these new insights, proactive and personalized strategies for enhancing safety were highlighted (Kim et al., 2016; Kamrani et al., 2017; Wali et al., 2018b). Methodologically, the importance of accounting for hierarchical heterogeneity in maximum-simulated and Bayesian framework was emphasized (Wali et al., 2018b; Arvin et al., 2019).

---

[2] Note that real-world driving data generated by connected vehicles, radar sensors, or video surveillance are typically used for quantifying the extent of variations in instantaneous driving decisions (Liu et al., 2015; Liu and Khattak, 2016; Khattak and Wali, 2017). Extreme driving behaviors (based on information in police crash reports) are generally referred to as "aggressive driving" in the literature (Nevarez et al., 2009; Paleti et al., 2010). However, we prefer to use a neutral term, driving volatility, to refer to the variations in real-world microscopic driving performance and the extreme (potentially unsafe) behaviors therein.

It seems reasonable to expect that the variations in microscopic driving behaviors *immediately* prior to crash involvement, termed as driving volatility, can be majorly correlated with crash outcomes, i.e., injury severity. The previous studies focused on crash frequency and not on the outcomes of crashes per se (Kim et al., 2016; Kamrani et al., 2017; Wali et al., 2018b). Also, previous studies analyzed aggregated data in the sense that location-specific behavioral data were linked with historic police-reported crashes at such locations. As such, insights regarding how driving volatility in time to collision relates to driver's propensity of receiving injuries cannot be obtained. With the recent advent of Naturalistic Driving Study (NDS) data, driving decisions in time to collision can be analyzed vis-à-vis driving behavior in normal events. To this end, recent studies have shown that the variations in microscopic driving decisions in time to collision can be a leading indicator of safety (Wali et al., 2019b; Wali and Khattak, 2020). Using the concept of "event-based volatility", the statistical relationships between instantaneous driving decisions in longitudinal and lateral directions and crash propensity (normal driving, crash, near-crash) are established (Wali et al., 2019b; Wali and Khattak, 2020). For example, greater *"intentional"* volatility in non-vulnerable and vulnerable locations (e.g., school zones) is reported to increase the likelihood of near-crashes/crashes (Wali et al., 2019b; Wali and Khattak, 2020). In line with the conceptual argument presented in the literature with relevance to the potential presence of heterogeneity in naturalistic driving data (see Table 1 in (Mannering et al., 2016)), both of the above studies concluded presence of substantial variations in the effects of naturalistic driving volatility on crash propensity due to systematic variations in factors unobserved in the data (Wali et al., 2019b; Wali and Khattak, 2020). Nonetheless, the previous studies did not focus on the injury severity component – i.e., little is known about how driving volatility in time to collision relates with injury outcomes.

**Research Objective and Contribution**

This study focuses on the links between driving volatility and crash-injury severity. A tight observational study design is harnessed to quantify real-world driving volatility in time to collision, and how it relates to injury outcomes sustained by drivers. In particular, the study uses a unique Naturalistic Driving database of drivers involved in crash events. Large-scale data on real world instantaneous driving behaviors in time to collision are analyzed to create driving volatility indices using different driving performance measures. To quantify the links between driving volatility in time to collision and crash-injury severity outcomes, the volatility indices are then combined with individual-level data on injury severity, event-specific variables including pre-crash behaviors, maneuvers, fault status, secondary tasks and durations, roadway, traffic, and environmental factors. In doing so, the critical methodological concerns related to observed (systematic) and unobserved (random) heterogeneity are carefully addressed (Mannering et al., 2016). From a methodological standpoint, fixed and random parameter discrete outcome logit models are developed to account for both systematic and random heterogeneity. Possible heterogeneity-in-means and variances of the random parameters varying as a function of several observed explanatory factors is captured. By using advanced modeling methods, the study contributes by seeking a fundamental understanding of short-term microscopic driving volatility, and how can we map driving volatility to injury severities sustained by drivers in crashes. An analysis along these lines is indispensable for developing more personalized behavioral countermeasures to potentially reduce drivers' injury outcomes.

# METHODOLOGY
## Conceptual Illustration

To understand the links between driving volatility in time to collision and injury outcomes (given a crash), detailed microscopic instantaneous driving data are needed. The SHRP2 NDS provides unique and relevant data (TRB, 2013). A key aspect of SHRP2 NDS is that it provides information on real-world driving decisions undertaken by drivers prior to involvement in a crash event. Crash is defined as "any contact that the subject instrumented vehicle has with an object (moving or fixed) at any speed in which kinetic energy is measurably transferred or dissipated" (Hankey et al., 2016). Figure 1 shows the conceptual framework for the analysis of instantaneous driving decisions in time to collision. Importantly, the instantaneous driving data are event specific (in our case event is a crash), and thus facilitate analysis of driving volatility and its correlation with injury outcomes while controlling for a wide variety of traffic, roadway, and behavioral factors in the event detail table (Figure 1).

The data from on-board data acquisition systems (DAS) can be used to characterize volatility in instantaneous driving decisions (Figure 1). In particular, the on-board DAS collect high-resolution (frequency of 10 Hz) movement related data (Hankey et al., 2016). Since instantaneous driving decisions immediately prior to involvement in unsafe events are more crucial, the SHRP2 NDS provides 30 seconds instantaneous driving behavior data for every crash event (Figure 1).

**(PLACE TABLE 1 ABOUT HERE)**

## Components of Volatility

A 30-seconds longitudinal acceleration profile in time to collision for a sampled event is shown in Figure 1. Harnessing such high-resolution motion data, driving volatility metrics can be created for each of the crash events using appropriate big data analytic techniques. When linked with unsafe outcomes, deeper insights can be obtained regarding the mechanisms through which variations in real-world driving behavior can influence crash-injury outcomes. We hypothesize a positive correlation between driving volatility and injury outcomes.

From a behavioral standpoint, considering the entire 30-seconds motion data aggregates different safety-relevant components of driving volatility. The aggressive driving behaviors when a driver is in control of the situation or vehicle can be termed as "intentional volatility", whereas, volatility arising due to evasive maneuvers or loss of control after a driver anticipated an unsafe event can be regarded as "unintentional volatility". Intentional volatility can be captured by analyzing instantaneous driving behavior data 20 to 30 seconds before the crash whereas the situational or unintentional volatility can be manifested through the motion data immediately prior to the crash (e.g., 10 seconds before). To this end, segmented volatility indices derived from using different durations of driving data are also created which can distinguish between intentional and unintentional volatility in time to collision (Figure 1).

## Calculation of Volatility

For creation of volatility measures in longitudinal direction, speed, acceleration, and/or steering angles based measures can be utilized (Quddus, 2013; Liu and Khattak, 2016). The present study uses acceleration/deceleration profiles for generation of volatility indices. Typically, the deceleration profiles exhibit larger variations (Wali et al., 2018b; Wali et al., 2019b; Wali and Khattak, 2020). Thus, separate volatility indices are created for capturing variations in acceleration and deceleration in the longitudinal direction. Statistically, coefficient of variation ($C_v$ or $CV$) is used as a measure for characterizing driving volatility in time to collision. The scale-insensitive property of $C_v$ is desired since it allows for a direct comparison of the longitudinal volatility in time to collision between different crash events (Wali et al., 2018b). Once acceleration and deceleration values are separated, $C_v$ is calculated for both by dividing the standard deviations of acceleration (deceleration) by the mean (average) values, i.e., $\sigma/\mu$. The process is repeated for all crash events, and both by using the entire 30-seconds data and bin-wise 10 seconds data to better characterize the complex mechanisms through which driving volatility in time to collision can influence crash-injury outcomes (Figure 1). In total, eight volatility measures are computed for longitudinal direction (shown later). Instantaneous driving decisions in the lateral dimension, such as lane change decisions, are also important as larger variations in lateral dimension may also correlate with crash-injury outcomes (Wali et al., 2019b; Wali and Khattak, 2020). Having said this, separate volatility indices (for acceleration and deceleration) are also generated for instantaneous driving decisions in the lateral dimension. This resulted in an additional eight volatility measures in the lateral dimension.

## Statistical Models

Once the 16 different volatilities indices are calculated for each crash event, the links between injury severity and driving volatility are modeled while controlling for different observed and unobserved factors. Past research has extensively used a variety of methodological alternatives for modeling crash-related injury severity including multinomial or binary probit/logit models, ordered choice models, nested logit models, and others (see Savolainen et al. (2011) and Mannering and Bhat (2014) for detailed reviews of methodological alternatives). A multinomial logit discrete outcome framework is used in the present study owing to the discrete nature of the response outcome. Following (McFadden, 1973; Shankar and Mannering, 1996), an injury severity function determining the outcome *k* of a specific event *n* can be defined as:

$$IS_{kn} = \boldsymbol{\beta_k} \boldsymbol{X_{kn}} + \varepsilon_{kn} \tag{1}$$

where $IS_{kn}$ is an injury severity function determining the injury outcome *k* for event *n*; $\boldsymbol{X_{kn}}$ is the vector of explanatory variables; $\boldsymbol{\beta_k}$ contains the estimable parameters related to each of the explanatory factor in $\boldsymbol{X_{kn}}$, and $\varepsilon_{kn}$ are the error terms. The multinomial logit model arises when a generalized extreme value distribution is assumed for $\varepsilon_{kn}$ (Shankar and Mannering, 1996):

$$P_n(k) = \frac{\exp[\boldsymbol{\beta_k} \boldsymbol{X_{kn}}]}{\sum_K \exp[\boldsymbol{\beta_k} \boldsymbol{X_{kn}}]} \tag{2}$$

where $P_n(k)$ is the probability of a specific outcome *k* (from the super-set of all possible injury categories *K*) for event *n*. To calculate estimates of $\boldsymbol{\beta_k}$, the following log-likelihood function is solved:

$$LL = \sum_{n=1}^{N} \left( \sum_{k=1}^{K} \aleph_{kn} \left[ \boldsymbol{\beta_k X_{kn}} - LN \sum_{\forall K} exp(\boldsymbol{\beta_k X_{Kn}}) \right] \right) \qquad (3)$$

where $\aleph_{kn}$ is an indicator equal to 1 if the observed injury outcome for event *n* is *k*, and 0 otherwise (Train, 2009). Following Train (2009), individual-level marginal effects averaged across the sampled events are reported since marginal effects can vary depending on the levels of independent variables (Train, 2009; Ahmad et al., 2019a).

### *Systematic (Observed) and Random (Unobserved) Heterogeneity*

The key objective of this study is to examine the links between driving volatility related measures and crash-injury severity outcomes. Some of the factors influencing crash-injury severity outcomes can be observed in the data while other factors (potentially important) can be missing in the data at hand. Data extracted from police-reported crash forms, traffic and local weather stations, and in some cases highway-asset-management systems have been historically used to understand the factors associated with injury severity outcomes (Mannering and Bhat, 2014). However, with the existing databases, there exists a real possibility that data on all the factors known to influence injury severity outcomes may not be available for analysis. For example, detailed information about driver-related behavioral variables is not generally available, and such variables may be correlated with crash-injury outcomes. Likewise, while safety-feature indicators (air-bags deployment, safety belts usage etc.) are typically available in traditional datasets, the effectiveness of these safety features in reducing crash-injury severity may vary across different crashes due to driver-specific characteristics such as height, health conditions, and bone density (to name a few), and such information is not typically available in traditional crash datasets (Mannering et al., 2016). These factors (potentially important) constitute what is referred to as "unobserved heterogeneity" in the safety literature (Mannering et al., 2016), and which is reflective of the possibility of systematic variations in the effects of explanatory factors across the sample population due to unobserved factors[3]. As explicitly noted in Mannering et al. (2016), emerging data sources such as naturalistic driving data provide additional valuable information but still may not be enough to fully model the factors correlated with crash-injury severity outcomes (Mannering et al., 2016). Such unobserved factors can potentially introduce heterogeneity in the effects of observed explanatory factors on crash-injury severity. Recall that the focus of the current study is to understand the relationship between crash-injury severity and driving volatility. The driving volatility indices are calculated based on the vehicle kinematics data collected by on-board units installed in vehicles participating in the naturalistic driving study. As noted in Mannering et al. (2016), the kinematics data are vehicle-specific (and driver-specific), and can vary significantly across different vehicles and drivers, and which can introduce heterogeneity in the effects of "observed" driving volatility-related variables on crash-injury severity. This is empirically demonstrated in recent studies focusing on the links between driving volatility and crash propensity in naturalistic driving environments (Wali et al., 2019b; Wali and Khattak, 2020). In addition, if important explanatory factors are omitted from the models, and appropriate methodological remedies not taken, the "observed" correlation between driving volatility in time to collision and crash-injury severity outcome may not be the "true" association between the two, but instead a manifestation of the effects of the omitted factors.

---

[3] For a detailed discussion on why unobserved heterogeneity may make the effects of explanatory factors vary across the sample, interested readers are referred to Mannering et al. (2016).

Given this important methodological concern, statistical methods that can account for unobserved heterogeneity in the crash-injury severity analysis have fairly become a methodological standard (Mannering and Bhat, 2014; Mannering et al., 2016). By allowing the effects of exogenous explanatory factors to vary across individual crashes (or segments of population), more efficient, precise, and richer insights can be obtained. To account for unobserved heterogeneity, a broad spectrum of studies have successfully used different methodological alternatives including random parameter models (Anastasopoulos and Mannering, 2009; Zhao and Khattak, 2015; Alarifi et al., 2017; Bhat et al., 2017; Khattak et al., 2019; Khattak and Fontaine, 2020), correlated random parameter models (Fountas et al., 2018a; Fountas et al., 2019; Wali et al., 2019a; Matsuo et al., 2020), random parameter models with heterogeneity-in-means (Venkataraman et al., 2014; Behnood and Mannering, 2017b; Wali et al., 2018c; Hamed and Al-Eideh, 2020), random parameter models with heterogeneity-in-means and variances (Behnood and Mannering, 2017a; Seraneeprakarn et al., 2017; Xin et al., 2017; Behnood and Mannering, 2019; Al-Bdairi et al., 2020; Yu et al., 2020), latent-class models (Eluru et al., 2012; Behnood et al., 2014; Shaheed and Gkritza, 2014; Yasmin et al., 2014a; Fountas et al., 2018b), latent class models with random parameters (Xiong and Mannering, 2013), Markov-switching models (Malyshkina and Mannering, 2009; Malyshkina et al., 2009; Khattak and Wali, 2017), Markov-switching models with random parameters (Xiong et al., 2014), and copula based approaches (Eluru et al., 2010; Yasmin et al., 2014b; Wang et al., 2015a; Wali et al., 2018a; Wali et al., 2018e; Wang et al., 2019). For a detailed discussion on the advantages and limitations of each of these methods, see Mannering et al. (2016). In the current study, we account for (possible) systematic variations in the effects of explanatory factors through random parameter modeling technique. To account for the unobserved heterogeneity in the discrete outcome probability process, random parameters can be introduced as (Milton et al., 2008; El-Basyouny and Sayed, 2009; El-Basyouny and Sayed, 2011; Anastasopoulos et al., 2012; Fountas and Anastasopoulos, 2017):

$$\beta_n = \beta + \Upsilon\zeta_n \tag{4}$$

where $\beta$ is the mean of random parameter vectors, $\mathbf{Y}$ is the diagonal matrix with standard deviations for random parameters, and $\zeta_n$ is a randomly distributed random term that captures unobserved heterogeneity across crashes (Tay, 2015; Mannering et al., 2016). In particular, the distribution for $\zeta_n$ is specified by the analyst where different distributions can be tested (discussed later). The estimation proceeds with simulated maximum likelihood procedures where 200 Halton draws (instead of random draws) are used to approximate the complex integrals over unobserved densities (Bhat, 2003). The results however did not change for Halton sequences as large as 1000. To explore the different possible contours related to unobserved heterogeneity, different functional forms are tested for the parameter density functions with normal distribution resulting in the best-fit. While we did not have a conceptual motivation for testing constrained distributions especially for driving volatility measures, for completeness – the superset of distributions tested included normal, log-normal, triangular, uniform, skewed and truncated normal, Weibull, beta, and exponential distributions. Further details can be found in (Bhat, 2003; Anastasopoulos and Mannering, 2009).

By estimating different set of crash-specific coefficient $\beta_n$, the mathematical exposition in Eq. 4 captures unobserved heterogeneity. However, the means of the random parameters are still held fixed across the observations. To account for the highly likely possibility of heterogeneity in the means and variances of random parameters, we model the location and shape parameters underlying the random parameters as a function of observed independent variables[4]. Thus, Eq. 4 becomes (Venkataraman et al., 2014; Behnood and Mannering, 2017b):

$$\beta_n = \boldsymbol{\beta} + \boldsymbol{\xi} \boldsymbol{Z_n} + \sigma_n exp(\beth_n \boldsymbol{B_n}) \boldsymbol{v}_{\,n} \tag{5}$$

where $\beta$ is the mean parameter estimate across all crashes $i$, $\boldsymbol{Z_n}$ is a vector of explanatory factors from crash $n$ which influence the mean of $\beta_n$, $\boldsymbol{\xi}$ is the parameter vector associated with $\boldsymbol{Z_n}$, $\boldsymbol{B_n}$ is a vector of explanatory factors that captures heterogeneity in the standard deviation of random parameter $(\boldsymbol{\sigma_n})$, $\beth_n$ is a parameter vector associated with $\boldsymbol{B_n}$, and $\boldsymbol{v_n}$ is the disturbance term. In addition to accounting for unobserved heterogeneity, the formulation in Eq. 5 now also accounts for observed heterogeneity. This is particularly useful as it allows to understand the factors that drive the contours of unobserved heterogeneity due to systematic variations in otherwise random factors. Importantly, from an estimation standpoint, unlike the common practice of using low number of Halton draws in specification search and then estimating the final model with a larger number to keep estimation cost under control, we used 200 Halton draws as a starting point to avoid poor approximation to the integral, model overspecification and missing statistically significant heterogeneity-in-variance terms that could otherwise arise with following the common practice. Compared to the common practice, the approach we used ensures that the support for parameters related to systematic and random heterogeneity is fully explored.

---

[4] Compared to the traditional random parameter (with fixed-means) models, Behnood and Mannering (2017b) found that accounting for heterogeneity-in-means in logit models resulted in better fit and substantially different inferences (Behnood and Mannering, 2017b). In addition, recent studies have extended the heterogeneity-in-means approach to also account for heterogeneity-in-variances (Xin et al., 2017; Behnood and Mannering, 2019; Al-Bdairi et al., 2020; Yu et al., 2020). The study by Seraneeprakam et al. (2017) found significant differences in the magnitudes of direct marginal effects for random parameters logit models with no mean-variance heterogeneity, with mean-only heterogeneity, and with both mean-variance heterogeneity (Seraneeprakarn et al., 2017). Likewise, recent studies have shown that constraining the means and variances of random parameters without statistical validation can result in model specification error, further leading to misguided policies (Xin et al., 2017; Behnood and Mannering, 2019; Al-Bdairi et al., 2020; Yu et al., 2020).

# DATA

The 2nd Strategic Highway Research Program (SHRP2) Naturalistic Driving Study (NDS) is the largest naturalistic driving environment till date (TRB, 2013), including 3,400 participant drivers with over 4,000 years of real-world naturalistic driving data collected between 2010 and 2013[5] (Hankey et al., 2016). In particular, this study uses the "event data" and "continuous" time-series data collected as part of NDS (Dingus et al., 2016; Wali et al., 2019b; Wali and Khattak, 2020). For crash-involved drivers, the event table provides comprehensive data on pre-crash driving behaviors, maneuvers (and its legality), start and end times of up to three secondary tasks (if applicable), passengers in front- and rear-seat, roadway, traffic, and environment related factors (Dingus et al., 2016; Wali et al., 2019b; Wali and Khattak, 2020). Secondary tasks are defined as any observable driver engagement other than the key driving tasks, and which may begin at any point during the 5 seconds prior to the event start, i.e., crash in this case, through the end of the event (TRB, 2013).

In this study, 671 crash events are analyzed in which 501 distinct drivers are involved, i.e., some participants had more than one crash during the study period. For the thousands of instrumented participant vehicles, DAS are used to collect video streams related to forward and rear roadway, driver's face view and hand status, integrated with information on vehicle network and status (speed, brake status, gas pedal, steering wheel position, acceleration), and data from additional DAS sensors (e.g., accelerometers) (TRB, 2013; Wali et al., 2019b; Wali and Khattak, 2020). As discussed earlier, 30 seconds of real-world motion data for crash events are generally available[6]. Over two million records of real-world driving packets are processed and finally around 200,000 (crash-related) instantaneous motion packets are used for calculation of 16 different volatility measures. The comprehensive event data are eventually integrated with the crash-specific volatility indices for subsequent analyses[7].

---

[5] Note that using three years of data could introduce issues related to potential temporal instability of the parameter estimates. Several recent studies have demonstrated the presence of temporal instability in parameter estimates over time (years), especially during economic down turns (Behnood and Mannering, 2015; Behnood and Mannering, 2016; Mannering, 2018; Alnawmasi and Mannering, 2019; Islam and Mannering, 2020). A step further, time-of-day and week effects are found to be temporally unstable as well (Behnood and Mannering, 2019; Behnood and Al-Bdairi, 2020). However, other research has found temporal stability (i.e., that parameters are stable over similar time horizons) when pooling data for three years or less (Dabbour et al., 2020). If temporal instability is present in the NDS data used in this study, the estimation results will characterize these temporal variations as unobserved heterogeneity. Thus, our forthcoming findings should be interpreted with some caution in that temporal instability could be affecting at least some of our findings related to heterogeneity.

## RESULTS

### Descriptive Statistics

Table 1 presents the descriptive statistics of key variables used in this study. In the SHRP2 NDS database, the crash severity is coded into four categories: low-risk tire strike, minor crash [MC], police-reportable crash [PRC], and most severe crash [SC]. For detailed definitions of the different response outcome categories, see Hankey et al. (2016). As shown in Table 1, approximately 40% and 38% of crashes resulted in low-risk tire strike and minor crash respectively. Whereas 13.3% and 8.8% of crashes were police-reportable and most severe crashes respectively (Table 1).

Descriptive statistics of aggregate volatility measures calculated using the entire 30-seconds pre-crash motion data are presented (Table 1). Both for volatility measures in longitudinal and lateral direction, and for acceleration and deceleration, the volatility distributions on-average are highly dispersed, as shown by the substantial standard deviations associated with each volatility measure. An interesting finding is that the volatility in lateral acceleration is greater than the volatility in longitudinal acceleration. This may reflect the evasive maneuvers (such as abrupt lane change) that drivers may undertake to avoid the obstacles once they anticipate a crash.

---

[6] Based on the data available to the authors, the duration of time-series kinematics data available for safety-critical events (crashes) is not fixed. That is, we have some cases with less than 15 seconds of data for crashes – 45 crash events (out of 671 crash events) have a time series data of less than 15 seconds. The methodology presented earlier for calculation of segmented (intentional) driving volatility measures accounts for this issue and to that end our estimates of intentional driving volatility measures (especially based on first 10 seconds of driving data) are conservative.

[7] It has been shown that relatively small datasets (which is the case in this study) could significantly influence the model performance. Fe and Lord (2014) examined the effects of sample size (ranging from 100 to 10,000 observations) on estimable parameters for crash severity modeling comparing ordered probit, multinomial logit, and mixed logit frameworks (Ye and Lord, 2014). Variations in point estimates of constant terms and injury specific variables were observed. Compared to low sample size (e.g., 300 observations), the confidence intervals for mean parameters (and scale parameters in the case of mixed logit model) narrowed with increase in sample size (Ye and Lord, 2014). The study recommended a sample size of 5,000 observations for injury severity modeling using a mixed logit approach (Ye and Lord, 2014). To this end, our forthcoming estimation results should be interpreted with some caution in that the point estimates and statistical significance may change when more naturalistic driving data are used in future work. Also, while the study by Ye and Lord (2014) did not focus on mixed logit models with heterogeneity-in-means and variances, the scale parameters associated with random parameters in this study could be rather conservative estimates. That is, Ye and Lord (2014) observed that the confidence intervals for mean and scale parameters narrowed with increase in sample size for mixed logit models. Assuming this finding is applicable to mixed logit model with heterogeneous means and variances as well, then the (statistically significant) confidence intervals for heterogeneous means and variance terms in this study are likely to get even more precise with the use of additional data in future.

Compared to aggregate volatility measures, the descriptive statistics for segmented volatility indices are next presented in Table 1. Such a segmentation approach can distinguish between the intentional and unintentional volatility related components – and how volatility in time to collision relates to crash-injury severity. The results offer valuable insights. First, the distributions of segmented volatility measures (estimated based on time to collision) are on-average similar. It seems that for the sampled crashes, drivers exhibited erratic or volatility behavior well in advance of the crash, such as 20-30 seconds prior to the crash. Building on the above discussion, this also implies that intentional vs. unintentional volatility is on-average similar in magnitude. *Thus, the key question is how intentional driving volatility in time to collision correlates with injury severity outcomes?*

Second, for all the three bin-wise volatility indices, volatility in longitudinal deceleration on-average is greater than volatility in longitudinal acceleration. Given a crash event, this suggests that drivers decisions during deceleration are on-average more volatile. This finding is in agreement with the literature (Kim et al., 2016; Kamrani et al., 2017; Wali et al., 2018b; Wali et al., 2019a; Wali and Khattak, 2020).

Third, similar to the aggregate volatility indices, the volatility in lateral acceleration is greater than the volatility in longitudinal acceleration for the bin-wise volatility measures too. This finding is intuitive and may be an outgrowth of the crash avoidance maneuvers undertaken by the drivers in the lateral direction.

The descriptive statistics of other variables are also presented. For the sampled crash events, the mean speed is approximately 32 kilometers per hour. Out of the  around 200,000 instantaneous motion packets (discussed earlier), around 9% (N = 18,077) included zero speeds, i.e., the vehicle was not in motion at a specific time instance. These observations were removed from the calculation of volatility indices described earlier. Note, however, that including the zero speeds did not significantly change the distributions of volatility measures (results not shown here for brevity). In 35.9% of the crashes, drivers did not engage in secondary tasks, whereas drivers were texting in 3.9% of the crashes. Importantly, durations of secondary tasks are also available. On average, drivers spent 3.58, 0.77, and 0.14 seconds on first, second, and third secondary task, respectively. In addition, drivers undertook safe and legal maneuver in 72% of the crashes, safe and illegal maneuver in 2.5% of the crashes, unsafe and illegal maneuver in 14.5% of the crashes, and unsafe but legal maneuvers in 10.4% of the crashes. For a detailed description of maneuver judgement related variables, see Hankey et al. (2016). As a measure of multicollinearity, the variance inflation factors (VIF) for all variables are less than 3 (not show due to space constraints) – showing lack of problematic multicollinearity.

**(PLACE TABLE 1 ABOUT HERE)**

## Modeling Results

Two different model specifications are presented: 1) the first specification models crash severity as a function of aggregate volatility indices and other factors, and 2) the second specification models crash severity as a function of segmented volatility indices and other factors. For ease of discussion, we will refer to the two specifications as specification 1 and 2. In all the models, the severity function for low-risk tire strike is considered as a base outcome – i.e., the severity function was set to zero for low-risk tire strike crashes. As such, the results are interpreted with respect to low-risk tire strike crashes. Based on the variable specifications in the estimation results presented next, a total of 648 and 629 observations have complete data on relevant variables under specification 1 and specification 2, respectively.

*Model Specification 1*

For specification 1, Table 2 presents the goodness-of-fit statistics for the fixed parameter logit, random parameter logit, random parameter logit with heterogeneity-in-means, and random parameter logit with heterogeneity-in-means and variances. Whereas Table 3 shows the estimation results of all the models under specification 1. All the models are derived from a systematic process to include most important variables (such as driving volatility related factors and others) on the basis of statistical significance, specification parsimony, and intuition. First, fixed parameter multinomial logit models are estimated constraining estimable parameters to be fixed across the crash events (Table 2). However, if unobserved heterogeneity is present in the data, the correlations obtained through fixed parameter models are likely to be biased and imprecise. To address this, random parameter logit models are estimated that allowed the estimable parameters to vary across different crash events (Table 2). A particular variable is treated as a random parameter if the parameter estimates exhibited statistically significant shape (variance) terms or exhibited statistically significant location (mean) as well as shape terms (Fountas and Anastasopoulos, 2017). In the earlier case, results from the likelihood ratio test and AIC statistics are examined to compare the model treating the specific variable (with only statistically significant shape/standard deviation term) as random parameter with the model treating the parameters on the same variable as fixed (Fountas and Anastasopoulos, 2017). With fixed $\boldsymbol{\beta}$'s and varying $\boldsymbol{Y}$, the random parameter model accounts for the systematic variations in the effects of variables across the sample population due to unobserved factors. Two variables are found to be normally distributed random parameters suggesting that their effects vary across the crash events (Table 3). Interestingly, these variables are related to the volatility measures: Coefficient of variation in longitudinal direction in the injury propensity function of minor crash [MC] and mean speed (km/h) in the injury severity function of police-reportable crash [PRC] (Table 3).

To also account for systematic heterogeneity, random parameter models with heterogeneity-in-means and with heterogeneity-in-means and variances are estimated. Importantly, in addition to accounting for random heterogeneity, the random parameter heterogeneity-in-means approach now also accounts for observed heterogeneity by allowing the means of random parameters to vary as a function of specific observed factors. Both random parameters produced substantial heterogeneity-in-means as well. For mean speed (km/h), a subject driver at fault decreased the mean making police-reportable crash (third injury severity level) a less likely outcome (Table 3). Likewise, a subject driver at fault indicator decreased the mean of coefficient of variation in longitudinal direction making minor crash a less likely outcome (Table 3). Overall, these findings show that a portion of heterogeneity in individual-level β estimates associated with random parameters can be explained by observed factors which otherwise would be deemed unexplainable heterogeneity in the relatively simplistic random parameter model with no heterogeneity-in-means. A step further, when the random parameter heterogeneity-in-means modeling framework was extended to also consider potential heterogeneity-in-variances, the variance of random parameter for mean speed (km/h) in severity function for PRC was explained by duration of secondary tasks. Duration of both first and second secondary tasks increased the variance of random parameters associated with mean speed – providing deeper insights regarding the interactions between instantaneous driving speed and surrogates of distraction[8].

## (PLACE TABLE 2 ABOUT HERE)

To justify the use of different models, goodness-of-fit measures such as likelihood ratio test, AIC, BIC, and McFadden $R^2$ are used. After accounting for the degrees of freedom, random parameter logit model clearly outperformed its fixed parameter counterpart (Table 2), as reflected in lower AIC and BIC values and higher McFadden $R^2$ for the random parameter logit model. Next, accounting for observed/systematic heterogeneity further resulted in better fit as shown by the remarkably improved goodness-of-fit statistics of random parameter heterogeneity-in-mean model (Table 2). In particular, compared to the random parameter logit model, the AIC and BIC statistics of random parameter logit with heterogeneity-in-means reduced by around 41 and 38 units respectively (showing remarkable improvement for the random parameter model with heterogeneity-in-means). Finally, accounting for both heterogeneity-in-means and variances led to even further improvement in model goodness of fit with lowest AIC and BIC statistics and relatively highest McFadden $R^2$ (see Table 2). All these findings demonstrate the significant potential of heterogeneity-based models that account for systematic and random heterogeneity in extracting richer insights from the data at hand. Finally, to help conceptualize the distribution effects of random-held parameters at the population level, key distributional statistics are provided in Table 5, whereas Table 6 presents the average direct marginal effects of best-fit random parameter heterogeneity-in-means and variances model.

---

[8] For random parameters with heterogeneous means, the overall conditional population mean of the random parameter after accounting for systematic heterogeneity due to heterogeneity-in-means, or an estimate of $E[\beta_n|E[Z_n]] \approx E_Z E[\beta_n|Z]$, can be calculated as: $\beta_{mean, random\ parameter} + \delta \times \xi$. Where: δ is the sample mean of the variable ($Z$) as a function of which the mean of the random parameter varies and $\xi$ is the parameter vector associated with $Z_N$ (i.e., heterogeneity-in-mean parameter estimate). The estimated parameters from Table 3 and the corresponding sample means from Table 1 can be inserted in the above formulation to estimate the population-level means of the random parameters. The population mean of random parameters on mean speed (km/h) (in the severity function of police-reportable crash) at the mean proportion of subject driver at fault (0.854) is approximately 0.064 + 0.85×(-0.171) = -0.081. Likewise, the population mean for coefficient of variation in longitudinal direction (in the severity function of minor crash) while accounting for systematic heterogeneity due to subject driver on fault indicator is 2.638 + 0.958 × -1.951 = 0.768. This means that the population mean of parameters on coefficient of variation in longitudinal direction computed at the mean proportion of subject driver on fault is positive – increasing on-average the likelihood of minor crash compared to low-risk tire strike crash. Contrarily, the negative population mean of -0.081 for parameters on mean speed (km/h) at the mean proportion of subject driver on fault reveals that the likelihood of police-reportable crash is lower compared to low-risk tire strike crashes. This finding is alarming because it suggests that the behaviors (in terms of speeding) of at-fault drivers may have lesser negative effects on the host driver – especially when previous literature shows more negative consequences of the behaviors of at-fault drivers on the safety of non-at-fault drivers (discussed later in results section).

*Model Specification 2*

Using the segmented driving volatility measures as inputs, the goodness-of-fit results for fixed and random parameter logit models under specification 2 are presented in Table 2. The main motivation behind using segmented volatility measures is to separate out the different components of volatility (intentional vs. unintentional), and which can shed light on how volatility in time to collision is related to crash severity.

<div align="center">**(PLACE TABLE 3 ABOUT HERE)**</div>

In the random parameter logit model with segmented volatility indices (Table 3), a total of two variables are found to be normally distributed random parameters suggesting that their effects vary across crash events. These variables are coefficient of variation of acceleration in longitudinal direction (1st bin data) and mean speed (km/h) (1st bin data) in the severity function of minor crash (Table 3). To conceptualize the heterogeneity in "direction of effects" of the random parameters, distributional statistics are provided in Table 5. When the random parameter logit model was extended to account for observed (systematic) heterogeneity, both random parameters revealed significant heterogeneity-in-means as well (see the estimation results for best-fit random parameter heterogeneity-in-means model in Table 4). For mean speed (km/h) (1st bin), the indicator variable for subject driver on fault decreased the mean making minor crash less likely (Table 4). Likewise, the indicator variable for no secondary tasks decreased the mean of random parameter for volatility in longitudinal acceleration (1st bin) – making minor crash less likely compared to low-risk tire strike crash[9] (Table 4). The random parameter for mean speed (km/h) (1st bin) in the severity function for minor crash also exhibited statistically significant heterogeneity-in-variance – with the variance explained by indicator variable for darkness (but lighted) (Table 4). Similar to the results obtained under specification 1, the random parameter model with heterogeneity-in-means and variances resulted in best fit[10] (see the goodness-of-fit statistics in lower panel of Table 2).

It is important to note that for all the models presented in Table 3 and 4, the statistically significant heterogeneity-in-means and variances underscores the importance of our model specification for the SHRP2 NDS data used. Interesting findings regarding the correlations between driving volatility (particularly regarding segmented volatility), speed, secondary tasks and durations, maneuver judgments, and crash severity outcomes are discussed next[11,12].

## (PLACE TABLE 4 ABOUT HERE)

## DISCUSSION

### Safety Effects of Driving Volatility

The results and findings discussed here refer to the random-parameter models with heterogeneity-in-means and variances given its relatively best fit (Table 3 and 4). However, in order to compare the performance of this model with that of commonly applied fixed and random parameter models (with no heterogeneity-in-means and variances), we also show the model estimation results for fixed parameter and mixed logit models in Table 3 and 4. Overall, statistically significant and positive correlations are observed between the four aggregate volatility measures and crash severity outcomes (Table 3). This suggests that greater driving volatility (both in longitudinal and lateral directions) during the 30-seconds prior to crash occurrence increases the likelihood of more severe crash outcomes compared to low-risk tire strike crashes. See the average (direct) marginal effects in Table 5 which measure the change in resulting probability of the corresponding injury severity outcome due to a unit change (or change from "0" to "1" for dummy variables) in the value of the specific independent variable[13] (Quddus et al., 2002; Ahmad et al., 2019b). For instance, an increase in driving volatility in longitudinal and lateral directions increased the likelihood of minor, police-reportable, and most severe crashes (Table 3). Importantly, compared to the effect of volatility in longitudinal acceleration on the likelihood of most severe crashes ($\beta = 1.013$), the effect of volatility in longitudinal deceleration is significantly greater in magnitude ($\beta = 4.826$). For instance, a unit increase in volatility in longitudinal acceleration on-average increases the probability of most severe crash by 0.0418 units – compared to a 0.2654 unit increase in the probability of most severe crash with a unit increase in volatility in longitudinal deceleration. Similar greater effects of volatility in longitudinal deceleration on the likelihood of police-reportable crashes [PRC] and minor crashes [MC] (compared to low-risk tire strikes) are observed (Table 3). These findings are important in that it suggest that while greater driving volatility is generally unsafe – a greater magnitude of volatility during braking could be more consequential. Note that, mean speed during the 30-seconds prior to crash is also included in the specification which intuitively suggests that higher speeds are associated with high order crash outcomes (Table 3). The parameter estimates for mean speed (km/h) [PRC severity function] and coefficient of variation in longitudinal direction [MC severity function] exhibited directional heterogeneity at the population level as well - with positive effects for 75.2% and 97.5% of the crashes respectively and negative for the rest[14] (see the heterogeneity distributions for the best fit heterogeneity-in-means and variances model in Table 4).

---

[9] The overall conditional population means for these random parameters while accounting for the systematic heterogeneity due to heterogeneity-in-means can be calculated using the procedure in footnote 8.

[10] Note that under both specifications extensive testing was done to consider all the explanatory variables listed in Table 1 as random parameters and for the possibility of heterogeneous effects in the means and variances of random parameters before arriving at the final specifications shown in Table 3 and 4. The substantial improvements in the goodness of fit statistics for heterogeneity-in-means and variances models reflect the careful testing conducted in this regard.

[11] As a crude rule-of-thumb for getting unbiased estimates, all the dummy variables included under specification 2 (Table 4) have at least 30 observations of "1s" except the indicator variable for cell phone texting which has 26 observations. This may lead to some bias in the parameter estimates for this indicator variable. However, this variable was retained in the model given its conceptual importance. In future, as more NDS data become available (such as those used in (Wali and Khattak, 2020)), this limitation can be addressed.

[12] Note that some of the variables in Table 4 are statistically insignificant at any reasonable level of confidence, e.g., coefficient of variation of lateral acceleration (calculated using 2nd 10-seconds data bin), signalized intersection in the severity function of most severe crashes, and darkness (but lighted) in the severity function of police-reportable crash. However, these variables are theoretically plausible to include in the model and are retained to show that the statistical significance of variables can decrease when observed and unobserved heterogeneity is adequately accounted at the same time. Also, removing these variables led to around 10 and 8 points increase in AIC and BIC, respectively (indicating poorer data fit). Note that AIC and BIC penalize adding more variables and discourages overfitting, so this reduction in AIC and BIC values is meaningful and seems substantial.

While the above findings regarding volatility and crash outcomes are interesting and new, the findings do not shed light on how volatility in time to collision is related to crash severity. This is important in the sense that if drivers' (intentional) volatility well in advance of a crash (20-30 seconds before the crash) is positively correlated with crash outcomes, control assists and warnings can be given to drivers in real-time to reduce unsafe and erratic driving behavior and decrease the likelihood of more severe crash

---

[13] To capture the variations in individual-level marginal effects arising due to the intrinsic nonlinearity in logit models, we also show the descriptive statistics of individual-level direct marginal effects for each of the independent variables.

[14] Importantly, ignoring observed heterogeneity in the means and variances of the random parameters (as is done in the typical mixed logit model) led to overestimation of the directional heterogeneity contours at the population level (see Table 4). As with all random parameter models, note that the effects regarding directional heterogeneity must be interpreted with caution. The finding that the associations between volatility and crash severity outcomes are negative in some cases does not imply causation. Instead, it suggests the role of other context-specific 'unobserved factors' which when combined with driving volatility measures leads to a negative correlation between driving volatility and crash severity outcomes.

outcomes. Having said this, the results presented in Table 4 offer important insights. Several volatility measures calculated based on the first and second 10-second bins of driving data (capturing intentional volatility) are statistically significantly and positively correlated with crash severity outcomes. The volatility measures related to longitudinal acceleration and deceleration are both positively correlated with the likelihoods of police-reportable and most severe crashes. A one-unit increase in the coefficient of variations of longitudinal acceleration and deceleration ($1^{st}$ bin) is associated with 0.0488 and 0.0521 units increase in the probability of most severe crashes, respectively (Table 6). However, in some cases, a unit increase in the volatilities associated with longitudinal acceleration and deceleration (calculated using $1^{st}$ 10-second bin data) led to as much as 0.4011 and 0.5011 increase in the probability of most severe crashes (see the marginal effects in Table 6). Likewise, greater volatilities in lateral acceleration and deceleration (calculated using $1^{st}$ bin) are also statistically significantly and positively correlated with the likelihood of most severe crashes (though for volatility in lateral deceleration, the association is significant only at 85% confidence level). Coming to the severity function for police-reportable crashes, volatilities in longitudinal acceleration and deceleration (calculated using $1^{st}$ 10-second data bin), volatility in longitudinal deceleration and lateral acceleration (calculated using $2^{nd}$ 10-second data bind) are statistically significant at 90% confidence level (except volatility in lateral acceleration based on $2^{nd}$ bin which is significant at 80% confidence level). Mean speed calculated using first 10-seconds data is also statistically significantly correlated with higher likelihoods of minor, police-reportable, and most severe crashes compared to low-risk tire strike crashes. Under both specifications, note that the magnitude of parameter estimates for several volatility related variables not only differ between fixed and random parameter counterparts but also between the random parameter models which account for observed and/or unobserved heterogeneity[15] (see

---

[15] We note that the new findings related to the links between driving volatility and injury severity, i.e., the increase in injury outcomes with increase in intentional driving volatility, should be interpreted with some caution. As discussed in Wali et al. (2019b), some level of volatility is natural and as an outgrowth of surrounding driving environments, e.g., driving in complex urban settings. Also, since drivers must respond to surrounding contexts by accelerating and decelerating all the time, zero volatility is not practically achievable on an average trip and does not necessarily imply safer driving. Having said this, the key question then becomes what could be a reasonable volatility or a threshold indicating high risk of more severe injuries? Answering this question fundamentally translates to fully addressing systematic heterogeneity arising due to non-linear and/or interactive effects. While the methodological framework presented herein indirectly accounts for systematic heterogeneity (captured through heterogeneous means and variances), it does not explicitly model the contours of systematic heterogeneity especially those arising due to potential nonlinear effects of driving volatility. This requires application of methods that can fully account for potentially complex nonlinear effects (Wali et al., 2019b). Keeping in mind the need for adequate characterization of the underlying relationship, innovative hybrid machine-learning based nonlinear models can be employed to develop context-specific volatility thresholds while simultaneously accounting for random heterogeneity. Along this line, we recommend that both systematic and random heterogeneity components are simultaneously analyzed in analysis of such a nature. This is crucial because without explicitly accounting for both, it is impossible to discern the true source (random or systematic) of heterogeneity (Pinjari and Bhat, 2006; Mannering et al., 2016; Wali et al., 2018d; Wali et al., 2019a).

Table 3 and 4).

Form a proactive safety perspective, the findings related to segmented volatility have important implications. Greater "intentional volatility" in time to collision makes more severe injury outcomes more likely to occur. To this end, warnings and alerts can be generated if real-time driver's decisions in longitudinal and lateral directions exhibit greater volatility (in particular well ahead of a crash). In particular, such customized alerts can be very useful especially when microscopic driving behaviors during deceleration regime are more volatile and that the effect of intentional volatility in deceleration on crash-injury severity outcomes is more severe.

<div align="center">(PLACE TABLE 5 ABOUT HERE)</div>

## Safety Effects of Secondary Task Durations, Driver hand status and Legality of Maneuvers

The results also quantify the association between secondary tasks and crash severity outcomes. Results from the best-fit random parameter heterogeneity-in-means and variance (under specification 2) suggest that texting on cell phones increases the likelihood of police-reportable crash (Table 4). Compared to drivers who were not texting, those who did had on-average a 0.92% (0.0092×100) higher chance of police-reportable crash (Table 6). In some cases, however, the increase in the likelihood of police-reportable crash was as much as 35.3% (see the maximum marginal effect for this variable in Table 6). Note that the fixed parameter counterpart underestimated the effect of texting on cell phone on the likelihood of police-reportable crash (see the β estimates for this variable in Table 4). Regarding driver's hand status, results reveal that if both hands are on the wheels, the likelihood of minor crash (compared to low-risk tire strike) decreases (Table 4). Contrarily, if no hands were on the wheel, the likelihood of most severe crash increased on-average by 0.68 percentage points (Table 4 and 6). Finally, if the driver engaged in an unsafe but legal maneuver, the likelihood of minor crash (compared to low-risk tire strike) increased on-average by 1.66 percentage points.

## Safety Effects of Locality and Other Factors

Several locality related factors such as open residential area, school zones, signalized intersections, number of through lanes, and crashes in mist or light rain are positively associated with crash severity outcomes. Compared to other locations, the likelihood of most severe crashes in school zones on-average increased by 0.65 percentage points. This finding is intuitive as interactions between vulnerable and motorized users are more likely in school zones and given a crash a more severe crash outcome is more probable (Wali and Khattak, 2020). Note that this variable was statistically insignificant in the fixed parameter counterpart (Table 4) – underscoring the importance of accounting for observed and unobserved heterogeneity in modeling naturalistic driving data. As expected, the likelihood of police-reportable crash increased on-average by 1.28% at signalized intersections. Another interesting finding relates to the fault-status of the driver. If the subject driver is at-fault, the likelihood of most severe crash (for the subject driver) decreases on-average by 4.61 percentage points. Although this finding is in line with past research that shows that the characteristics (driving errors) of at-fault driver have more negative effects on injury severity of the not-at-fault driver (Russo et al., 2014; Wali et al., 2018e), this requires further investigation in future by simultaneously analyzing the crash outcomes of not-at-fault and at-fault driver in the context of the current study (Russo et al., 2014; Wali et al., 2018e).

## LIMITATIONS/FUTURE WORK

The present study is based on a sample of ~ 9800 events (baseline, near-crash, and crash events), out of which 671 were identified as crash events. However, the SHRP2 NDS Event Detail Table (EDT) currently has 41,479 records, out of which 1,877 are crash events (Wali and Khattak, 2020) (https://insight.shrp2nds.us/data/index). The authors used a subset of EDT due to lack of access to the entire SHRP2 NDS database. With regard to future work, there are several pathways for extending the proposed framework. As more data become publicly available, the methodology presented in this study can be expanded. Another extension of the research can be to apply the proposed methodology to specific roadway types.

## CONCLUSIONS

As a key indicator of unsafe driving, driving volatility characterizes the variations in instantaneous driving decisions, especially capturing extreme driving behaviors. This study characterized longitudinal and lateral volatility in microscopic driving decisions and examined how driving volatility in time to collision relates to crash-injury severity. A rigorous data analytic methodology was proposed to extract critical information embedded in real-world naturalistic driving data related to 671 crash events featuring around 0.2 million temporal driving samples. For the sampled 671 crash events, large-scale microscopic driving data in time to collision are analyzed to create volatility indices derived from different measures of driving performance. Aggregate and segmented volatility measures were created to adequately analyze intentional vs unintentional volatility. For the empirical analysis, the volatility indices are then linked with individual crash events including data on crash severity, event-specific variables such as drivers' pre-crash maneuvers and behaviors, traffic flow factors, secondary tasks and durations, roadway factors, and fault status. Separate crash severity outcome models are presented using aggregated and segmented volatility measures.

Overall, statistically significant positive correlations are found between the four aggregate volatility measures and crash severity outcomes. This suggests that greater driving volatility (both in longitudinal and lateral direction) during 30-seconds prior to crash occurrence increases the likelihood of more severe crash outcomes. Importantly, compared to the effect of volatility in longitudinal acceleration on crash outcomes, the effect of volatility in longitudinal deceleration is significantly greater in magnitude. Compared to the aggregate volatility measures, the results obtained from models with segmented volatility indices offered important insights. Several volatility measures calculated based on the first and second 10-second bins of driving data (capturing intentional volatility) were statistically significantly and positively correlated with crash severity outcomes. This highlights that greater driving volatility well before the driver anticipates a crash event (intentional volatility) increases the likelihood of most severe crash outcomes. Other interesting findings were discussed in detail. Form a proactive  safety perspective, the findings related to segmented volatility have important implications. By monitoring the instantaneous driving decisions in real-time, proactive alerts and customized warnings can be generated in case driving decisions exhibit greater volatility – potentially reducing the risk of more severe injuries.

From an empirical perspective, the study contributes by developing fixed- and random-parameter (with heterogeneity-in-means and variances) discrete outcome models that account for both systematic and random heterogeneity. Compared to the relatively simplistic random parameter models (with homogeneous means and variances), the models also account for observed heterogeneity-in-means and variances of the random parameters varying as a function of independent variables. In addition to offering deeper and more accurate insights, accounting for observed and unobserved heterogeneity led to substantial improvements in model goodness of fit.

## ACKNOWLEDGEMENT

The data for this study were provided through a collaborative effort between Virginia Tech Transportation Institute, the U.S. Federal Highway Administration (FHWA), and Oak Ridge National Laboratory (ORNL). The timely assistance and guidance of the ORNL team about data elements is highly appreciated. The authors would also like to recognize the contribution of Alexandra Boggs in proof-reading the manuscript. This paper is based upon work supported by the US National Science Foundation under grant No. 1538139. Additional support was provided by the US Department of Transportation through the Collaborative Sciences Center for Road Safety, a consortium led by The University of North Carolina at Chapel Hill in partnership with The University of Tennessee. Any opinions, findings, and conclusions or recommendations expressed in this paper are those of the authors and do not necessarily reflect the views of the sponsors. An earlier version of this paper was presented in a poster session in Transportation Research Board 97[th] Annual Meeting.

Wali, B., Khattak, A., Karnowski, T., 2020. The relationship between driving volatility in time to collision and crash-injury severity in a naturalistic driving environment. *Analytic Methods in Accident Research*, 100136.

Wali, B., Khattak, A., Karnowski, T., 2020. The relationship between driving volatility in time to collision and crash-injury severity in a naturalistic driving environment. *Analytic Methods in Accident Research*, 100136.

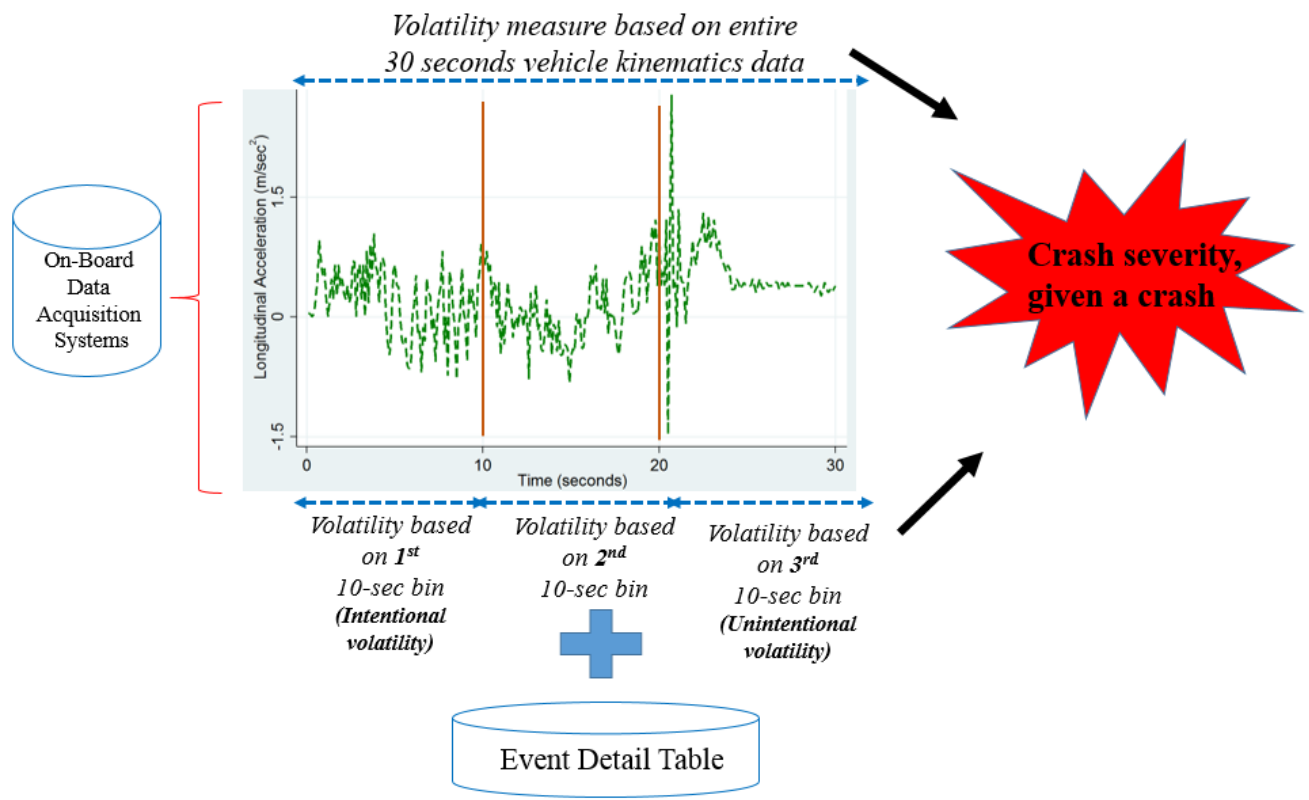

**FIGURE 1: Conceptual Framework**
(**Note:** The original data resolution is 10 Hz, i.e., one-tenth of a second. However, the x-axis is labelled in seconds (1 Hz) for ease of presentation.)

**TABLE 1: Descriptive Statistics of Key Variables**

| Category | Variable Name | N | Mean | SD | Min | Max |
|---|---|---|---|---|---|---|
| **Dependent Variable: Crash severity** | Low-risk tire strike | 671 | 0.404 | 0.491 | 0 | 1 |
| | Minor crash [MC] | 671 | 0.375 | 0.484 | 0 | 1 |
| | Police-reportable crash [PRC] | 671 | 0.132 | 0.339 | 0 | 1 |
| | Most severe [SC] | 671 | 0.087 | 0.283 | 0 | 1 |
| **Volatility based on entire 30-seconds driving data** | CV: longitudinal acceleration | 668 | 0.958 | 0.483 | 0.121 | 4.552 |
| | CV: longitudinal deceleration | 668 | 1.162 | 0.601 | 0.075 | 6.003 |
| | CV: lateral acceleration | 665 | 1.335 | 0.582 | 0.247 | 5.920 |
| | CV: lateral deceleration | 666 | 1.161 | 0.517 | 0.156 | 3.535 |
| | Mean Speed (km/h) | 668 | 31.629 | 21.508 | 0.527 | 121.175 |
| **Volatility based on first 10-seconds bin (K = 1)** | $CV\!: longitudinal\ acceleration_{K=1}$ | 665 | 0.902 | 0.401 | 0.224 | 4.169 |
| | $CV\!: longitudinal\ deceleration_{K=1}$ | 666 | 1.095 | 0.527 | 0.028 | 5.490 |
| | $CV\!: lateral\ acceleration_{K=1}$ | 661 | 1.253 | 0.563 | 0 | 5.736 |
| | $CV\!: lateral\ deceleration_{K=1}$ | 659 | 1.080 | 0.491 | 0.182 | 3.300 |
| **Volatility based on second 10-seconds bin (K = 2)** | $CV\!: longitudinal\ acceleration_{K=2}$ | 661 | 0.912 | 0.432 | 0.033 | 3.396 |
| | $CV\!: longitudinal\ deceleration_{K=2}$ | 662 | 1.079 | 0.528 | 0.063 | 3.998 |
| | $CV\!: lateral\ acceleration_{K=2}$ | 655 | 1.236 | 0.508 | 0.317 | 4.007 |
| | $CV\!: lateral\ deceleration_{K=2}$ | 661 | 1.044 | 0.479 | 0 | 3.739 |
| **Volatility based on third 10-seconds bin (K = 3)** | $CV\!: longitudinal\ acceleration_{K=3}$ | 643 | 0.903 | 0.406 | 0.115 | 3.672 |
| | $CV\!: longitudinal\ deceleration_{K=3}$ | 648 | 1.095 | 0.555 | 0.113 | 5.936 |
| | $CV\!: lateral\ acceleration_{K=3}$ | 642 | 1.228 | 0.496 | 0.157 | 4.493 |
| | $CV\!: lateral\ deceleration_{K=3}$ | 644 | 1.047 | 0.448 | 0.143 | 3.478 |
| **Passengers** | Number of front seat passengers (including driver) | 671 | 1.250 | 0.433 | 1 | 2 |
| | Number of rear seat passengers | 671 | 0.103 | 0.430 | 0 | 3 |
| **Travel lanes** | Number of through lanes | 671 | 1.213 | 0.938 | 0 | 5 |
| | Number of contiguous travel lanes | 671 | 2.293 | 1.575 | 0 | 7 |
| **Secondary tasks** | Holding cell phone | 671 | 0.028 | 0.166 | 0 | 1 |
| | Talking on cell phone: hand-held | 671 | 0.033 | 0.178 | 0 | 1 |
| | Texting on cell phone | 671 | 0.039 | 0.193 | 0 | 1 |
| | No secondary task | 671 | 0.359 | 0.480 | 0 | 1 |
| | Other tasks | 671 | 0.541 | 0.498 | 0 | 1 |
| **Duration of secondary tasks** | Duration in seconds of first secondary task | 671 | 3.585 | 4.190 | 0 | 24.119 |
| | Duration in seconds of second secondary task | 671 | 0.772 | 2.030 | 0 | 14.221 |
| | Duration in seconds of third secondary task | 671 | 0.145 | 1.044 | 0 | 13.878 |
| **Incident maneuvers** | Changing lanes | 671 | 0.031 | 0.174 | 0 | 1 |
| | Negotiating a curve | 671 | 0.075 | 0.263 | 0 | 1 |
| | Other maneuvers | 671 | 0.894 | 0.307 | 0 | 1 |
| **Maneuver judgement** | Maneuver is safe and legal | 671 | 0.723 | 0.448 | 0 | 1 |
| | Maneuver is safe and illegal | 671 | 0.025 | 0.157 | 0 | 1 |
| | Maneuver is unsafe and illegal | 671 | 0.145 | 0.352 | 0 | 1 |
| | Maneuver is unsafe but legal | 671 | 0.104 | 0.306 | 0 | 1 |

Notes: CV is coefficient of variation; N is sample size; SD is standard deviation; km/h is kilometers per hour.

**TABLE 1: Descriptive Statistics of Key Variables** (*Continued*)

| Category | Variable Name | N | Mean | SD | Min | Max |
|---|---|---|---|---|---|---|
| *Nature of events* | Conflict with a following vehicle | 671 | 0.054 | 0.225 | 0 | 1 |
| | Conflict with lead vehicle | 671 | 0.098 | 0.298 | 0 | 1 |
| | Other events | 671 | 0.847 | 0.359 | 0 | 1 |
| *Driver Behavior* | Exceeded safe speed but not speed limit | 671 | 0.054 | 0.225 | 0 | 1 |
| | Exceeded speed limit | 671 | 0.037 | 0.189 | 0 | 1 |
| | Distracted | 671 | 0.311 | 0.463 | 0 | 1 |
| | Made turn, cut corner on right | 671 | 0.146 | 0.353 | 0 | 1 |
| | Other behaviors | 671 | 0.451 | 0.498 | 0 | 1 |
| *Roadway factors* | Intersection influence: Traffic Signal | 671 | 0.185 | 0.388 | 0 | 1 |
| | Intersection influence: Uncontrolled | 671 | 0.083 | 0.277 | 0 | 1 |
| | Intersection influence: Stop sign | 671 | 0.063 | 0.242 | 0 | 1 |
| | Divided Roadway | 671 | 0.219 | 0.414 | 0 | 1 |
| | Not Divided - 2-way traffic | 671 | 0.484 | 0.500 | 0 | 1 |
| *Traffic factors* | Level of Service: A1 (Free flow, no lead traffic) | 671 | 0.562 | 0.496 | 0 | 1 |
| | Level of Service: A2 (Free flow, leading traffic present) | 671 | 0.180 | 0.385 | 0 | 1 |
| | Level of Service: B (Flow with some restrictions) | 671 | 0.180 | 0.385 | 0 | 1 |
| | Level of Service: Stable flow, maneuverability & speed more restricted | 671 | 0.049 | 0.216 | 0 | 1 |
| *Driver hand status* | Both hands on wheels | 671 | 0.465 | 0.499 | 0 | 1 |
| | Left hand only | 671 | 0.325 | 0.468 | 0 | 1 |
| | Right hand only | 671 | 0.143 | 0.350 | 0 | 1 |
| | None | 671 | 0.067 | 0.193 | 0 | 1 |
| *Seat-belt use* | Lap/shoulder belt properly worn | 671 | 0.900 | 0.300 | 0 | 1 |
| | None used | 671 | 0.095 | 0.279 | 0 | 1 |
| *Light conditions* | Darkness, lighted | 671 | 0.204 | 0.403 | 0 | 1 |
| | Darkness, not lighted | 671 | 0.046 | 0.210 | 0 | 1 |
| | Daylight | 671 | 0.708 | 0.455 | 0 | 1 |
| | Dawn/Dusk | 671 | 0.042 | 0.200 | 0 | 1 |
| *Weather factors* | Mist/Light rain | 671 | 0.058 | 0.234 | 0 | 1 |
| | No adverse weather | 671 | 0.860 | 0.347 | 0 | 1 |
| | Heavy rain | 671 | 0.061 | 0.240 | 0 | 1 |
| | Fog/snow | 671 | 0.021 | 0.143 | 0 | 1 |
| *Locality* | Business/industrial | 671 | 0.463 | 0.499 | 0 | 1 |
| | Moderate residential | 671 | 0.204 | 0.403 | 0 | 1 |
| | Open residential | 671 | 0.049 | 0.216 | 0 | 1 |
| | School | 671 | 0.079 | 0.270 | 0 | 1 |
| | Urban | 671 | 0.079 | 0.270 | 0 | 1 |
| | Other | 671 | 0.125 | 0.331 | 0 | 1 |
| *Fault status* | Other driver (Driver 2) on fault | 671 | 0.088 | 0.283 | 0 | 1 |
| | Subject driver on fault | 671 | 0.854 | 0.353 | 0 | 1 |
| | Other/not available | 671 | 0.058 | 0.234 | 0 | 1 |

Notes: N is sample size; SD is standard deviation.

**TABLE 2: Goodness of Fit Statistics for Alternative Heterogeneity-Based Model Specifications**

|  | Variable | Fixed parameter logit model | Random parameter logit model | Random parameter logit model with heterogeneity-in-means | Random parameter logit model with heterogeneity-in-means and variances |
|---|---|---|---|---|---|
| **First Specification** | N | 648 | 648 | 648 | 648 |
|  | No. of parameters | 31 | 33 | 35 | 37 |
|  | Log-likelihood at constant | -786.82 | -786.82 | -786.82 | -786.82 |
|  | Log-likelihood at convergence | -596.6 | -592.064 | -569.36 | -563.35 |
|  | McFadden Pseudo $R^2$ | 0.2295 | 0.3409 | 0.3661 | 0.3728 |
|  | Akaike Information Criterion | 1,255.2 | 1,250.1 | 1,208.7 | 1,200.7 |
|  | Bayesian Information Criterion | 1,280.36 | 1,276.91 | 1,237.13 | 1,230.73 |
| **Second Specification** | N | 629 | 629 | 629 | 629 |
|  | No. of parameters | 30 | 32 | 34 | 35 |
|  | Log-likelihood at constant | -786.82 | -786.82 | -786.82 | -786.82 |
|  | Log-likelihood at convergence | -600.65 | -594.22 | -584.04 | -580.54 |
|  | McFadden Pseudo $R^2$ | 0.1987 | 0.3185 | 0.3302 | 0.3342 |
|  | Akaike Information Criterion | 1,261.3 | 1,252.4 | 1,236.1 | 1,231.1 |
|  | Bayesian Information Criterion | 1,285.26 | 1,278.00 | 1,263.23 | 1,259.033 |

Notes: N is estimation sample size; First specification includes aggregate driving volatility measures as explanatory factors; Second specification includes segmented (intentional) driving volatility measures as explanatory factors.

**TABLE 3: Model Estimation Results for Crash Severity in Naturalistic Driving Environment (First-Specification)**

| Variable Name | Fixed Parameter Multinomial Logit Model | | Random Parameter Multinomial Logit Model | | Random Parameter Multinomial Logit Model - Heterogeneity-in-Means | | Random Parameter Multinomial Logit Model - Heterogeneity-in-Means and Variances | |
|---|---|---|---|---|---|---|---|---|
| | β | t-stat | β | t-stat | β | t-stat | β | t-stat |
| Constant [MC] | -3.102 | -6.38 | -3.322 | -5.58 | -3.223 | -5.29 | -3.329 | -5.14 |
| Constant [PRC] | -6.968 | -10.26 | -8.002 | -7.15 | -8.517 | -6.41 | -8.004 | -5.33 |
| Constant [SC] | -11.629 | -9.92 | -12.615 | -8.51 | -11.181 | -6.75 | -11.268 | -7.31 |
| **Volatility based on entire 30-seconds driving data** | | | | | | | | |
| CV: longitudinal acceleration [SC] | 0.973 | 2.80 | 1.089 | 2.710 | 0.984 | 2.19 | 1.013 | 1.84 |
| CV: longitudinal deceleration [SC] | 3.937 | 9.42 | 4.442 | 8.280 | 4.863 | 7.87 | 4.826 | 8.15 |
| CV: lateral acceleration [SC] | 0.948 | 2.80 | 0.994 | 2.61 | 1.153 | 2.71 | 1.106 | 2.23 |
| CV: lateral deceleration [SC] | 0.471 | 1.42 | 0.551 | 1.43 | 0.544 | 1.19 | 0.531 | 0.97 |
| Mean Speed (km/h) [SC] | 0.047 | 5.03 | 0.049 | 4.48 | 0.051 | 4.04 | 0.053 | 3.98 |
| CV: longitudinal acceleration [PRC] | 0.786 | 2.52 | 0.951 | 2.39 | 0.884 | 1.64 | 0.792 | 1.72 |
| CV: longitudinal deceleration [PRC] | 2.843 | 8.15 | 3.395 | 6.29 | 4.155 | 6.33 | 3.948 | 5.74 |
| CV: lateral acceleration [PRC] | 0.508 | 1.89 | 0.693 | 1.85 | 0.836 | 1.93 | 0.812 | 1.89 |
| Mean Speed (km/h) [PRC] | 0.036 | 5.07 | 0.009 | 0.32 | 0.061 | 2.03 | 0.064 | 2.39 |
| standard deviation (normally distributed) | --- | --- | 0.054 | 1.51 | 0.142 | 2.57 | 0.094 | 2.31 |
| CV: longitudinal acceleration [MC] | 0.551 | 2.15 | 0.614 | 1.97 | 0.522 | 1.54 | 0.541 | 1.31 |
| CV: longitudinal deceleration [MC] | 1.165 | 3.99 | 0.871 | 1.92 | 2.743 | 4.33 | 2.638 | 3.65 |
| standard deviation (normally distributed) | --- | --- | 1.185 | 2.25 | 1.169 | 2.37 | 1.344 | 2.41 |
| CV: lateral acceleration [MC] | 0.403 | 1.99 | 0.519 | 1.97 | 0.471 | 1.77 | 0.486 | 1.89 |
| Mean Speed (km/h) [MC] | 0.024 | 4.27 | 0.028 | 3.8 | 0.027 | 3.34 | 0.029 | 3.32 |
| **Secondary Tasks and Durations** | | | | | | | | |
| Cell phone, Texting [PRC] | 0.996 | 2.06 | 1.452 | 1.79 | 2.445 | 1.83 | 2.141 | 1.46 |
| Duration in seconds of 1st secondary task [MC] | 0.048 | 2.29 | 0.075 | 2.49 | 0.086 | 2.85 | 0.098 | 2.92 |
| Duration in seconds of 2nd secondary task [MC] | 0.081 | 1.83 | 0.146 | 2.17 | 0.158 | 2.29 | 0.197 | 2.87 |
| **Driver Hand Status and Maneuver Judgement** | | | | | | | | |
| Both hands on wheel [MC] | -0.547 | -2.96 | -0.738 | -2.77 | -0.697 | -2.62 | -0.713 | -2.46 |
| No hands on wheel [SC] | 1.057 | 1.17 | 1.682 | 1.64 | 1.774 | 1.61 | 1.865 | 1.33 |
| Unsafe but legal maneuver [MC] | 0.618 | 2.13 | 0.826 | 2.01 | 0.888 | 2.12 | 0.952 | 2.09 |
| **Locality and Environmental Factors** | | | | | | | | |
| Open residential area [MC] | 1.194 | 2.74 | 1.758 | 2.67 | 1.739 | 2.45 | 1.718 | 2.21 |
| School zone [SC] | 0.773 | 1.19 | 0.918 | 1.19 | 0.924 | 1.06 | 1.038 | 0.97 |
| Intersection influence: Traffic Signal [PRC] | 0.656 | 2.10 | 0.892 | 1.99 | 0.811 | 1.21 | 0.844 | 1.31 |
| Intersection influence: Traffic Signal [SC] | 0.735 | 1.68 | 0.796 | 1.57 | 0.815 | 1.47 | 0.809 | 1.28 |
| Darkness, lighted [PRC] | -0.632 | -1.75 | -0.814 | -1.52 | -1.166 | -1.48 | -1.123 | -1.27 |
| Darkness, lighted [SC] | -1.695 | -2.61 | -1.887 | -2.52 | -2.229 | -2.75 | -2.277 | -2.81 |
| Number of through lanes [SC] | 0.721 | 3.66 | 0.854 | 3.51 | 0.842 | 3.01 | 0.863 | 2.66 |
| Mist/light rain [SC] | 1.739 | 2.83 | 1.981 | 2.72 | 1.855 | 2.29 | 1.861 | 1.81 |

Notes: CV is coefficient of variation; MC is minor crash; PRC is police-reportable crash; SC is most severe crash; (---) is Not Applicable.

**TABLE 3: Model Estimation Results for Crash Severity in Naturalistic Driving Environment (First-Specification) (Continued)**

| Variable Name | Fixed Parameter Multinomial Logit Model | | Random Parameter Multinomial Logit Model | | Random Parameter Multinomial Logit Model - Heterogeneity-in-Means | | Random Parameter Multinomial Logit Model - Heterogeneity-in-Means and Variances | |
|---|---|---|---|---|---|---|---|---|
| | β | t-stat | β | t-stat | β | t-stat | β | t-stat |
| **Fault Status** | | | | | | | | |
| Subject driver on fault [SC] | -1.099 | -2.80 | -1.543 | -2.93 | -3.952 | -5.11 | -3.882 | -4.05 |
| **Heterogeneity-in-means** | | | | | | | | |
| CV: longitudinal deceleration [MC]: Subject driver at fault | --- | --- | --- | --- | -1.961 | -3.59 | -1.951 | -2.71 |
| Mean Speed (km/h) [PRC]: Subject driver at fault | --- | --- | --- | --- | -0.181 | -2.86 | -0.171 | -3.11 |
| **Heterogeneity-in-means and variances** | | | | | | | | |
| Mean Speed (km/h) [PRC]: Duration of 1st secondary task | --- | --- | --- | --- | --- | --- | 0.046 | 1.94 |
| Mean Speed (km/h) [PRC]: Duration of 2nd secondary task | --- | --- | --- | --- | --- | --- | 0.085 | 1.71 |

Notes: CV is coefficient of variation; MC is minor crash; PRC is police-reportable crash; SC is most severe crash; (---) is Not Applicable.

**TABLE 4: Model Estimation Results for Crash Severity in Naturalistic Driving Environment (Second Specification)**

| Variable Name | Fixed Parameter Multinomial Logit Model | | Random Parameter Multinomial Logit Model | | Random Parameter Multinomial Logit Model - Heterogeneity-in-Means | | Random Parameter Multinomial Logit Model - Heterogeneity-in-Means and Variances | |
|---|---|---|---|---|---|---|---|---|
| | β | t-stat | β | t-stat | β | t-stat | β | t-stat |
| Constant [MC] | -1.588 | -4.13 | -2.022 | -2.72 | -1.792 | -2.54 | -1.681 | -2.46 |
| Constant [PRC] | -5.647 | -9.00 | -6.451 | -8.86 | -6.485 | -8.94 | -6.551 | -8.92 |
| Constant [SC] | -10.255 | -9.14 | -11.285 | -8.71 | -11.212 | -8.62 | -11.205 | -8.34 |
| **Volatility Based on Segmented Driving Data** | | | | | | | | |
| CV: longitudinal acceleration, K=1 [SC] | 1.075 | 2.49 | 1.021 | 2.16 | 1.007 | 2.12 | 1.031 | 1.89 |
| CV: longitudinal acceleration, K=1 [SC] | 0.612 | 1.44 | 0.704 | 1.37 | 0.697 | 1.36 | 0.778 | 1.88 |
| CV: lateral acceleration, K = 1 [SC] | 0.791 | 2.77 | 0.827 | 2.59 | 0.845 | 2.63 | 0.835 | 2.40 |
| CV: lateral deceleration, K = 1 [SC] | 0.737 | 2.15 | 0.701 | 1.91 | 0.698 | 1.93 | 0.685 | 1.57 |
| Mean Speed (km/h), K = 1 [SC] | 0.044 | 4.84 | 0.051 | 4.82 | 0.052 | 4.86 | 0.051 | 4.45 |
| CV: lateral deceleration, K = 2 [SC] | 2.301 | 5.43 | 2.932 | 5.44 | 2.954 | 5.51 | 2.885 | 4.99 |
| CV: longitudinal acceleration, K=1 [PRC] | 0.705 | 1.83 | 0.631 | 1.63 | 0.619 | 1.59 | 0.624 | 1.87 |
| CV: longitudinal deceleration, K=1 [PRC] | 0.772 | 2.33 | 0.934 | 2.25 | 0.922 | 2.23 | 0.992 | 2.62 |
| CV: longitudinal deceleration, K = 2 [PRC] | 1.439 | 4.16 | 1.908 | 4.29 | 1.942 | 4.41 | 1.892 | 4.29 |
| CV: lateral acceleration, K = 2 [PRC] | 0.253 | 1.01 | 0.371 | 1.32 | 0.383 | 1.38 | 0.395 | 1.42 |
| Mean Speed (km/h), K = 1 [PRC] | 0.032 | 4.57 | 0.035 | 4.44 | 0.035 | 4.49 | 0.035 | 3.89 |
| CV: longitudinal acceleration, K=1 [MC] | 0.475 | 1.63 | 0.171 | 0.25 | 0.889 | 1.46 | 0.894 | 1.45 |
| standard deviation (normally distributed) | --- | --- | 2.721 | 2.32 | 2.251 | 2.14 | 2.002 | 1.99 |
| CV: lateral deceleration, K = 1 [MC] | 0.489 | 2.54 | 0.789 | 2.00 | 0.601 | 1.68 | 0.631 | 1.82 |
| Mean Speed (km/h), K = 1 [MC] | 0.025 | 4.56 | 0.029 | 2.49 | 0.041 | 2.81 | 0.039 | 2.76 |
| standard deviation (normally distributed) | --- | --- | 0.043 | 1.83 | 0.046 | 1.8 | 0.033 | 1.91 |
| **Secondary Tasks** | | | | | | | | |
| Cell phone, Texting [PRC] | 1.191 | 2.46 | 1.609 | 2.55 | 1.645 | 2.61 | 1.561 | 2.36 |
| **Driver Hand Status and Maneuver Judgement** | | | | | | | | |
| Both hands on wheel [MC] | -0.681 | -3.77 | -1.345 | -2.65 | -0.903 | -2.28 | -0.967 | -2.41 |
| No hands on wheel [SC] | 1.329 | 1.48 | 2.597 | 2.38 | 2.563 | 2.38 | 2.528 | 2.18 |
| Unsafe but legal maneuver [MC] | 0.659 | 2.28 | 1.543 | 2.03 | 1.686 | 2.32 | 1.473 | 2.24 |
| **Locality and Environmental Factors** | | | | | | | | |
| Open residential area [MC] | 1.211 | 2.74 | 2.868 | 2.20 | 2.508 | 2.09 | 2.658 | 2.24 |
| School zone [SC] | 0.803 | 1.27 | 1.191 | 1.67 | 1.201 | 1.68 | 1.292 | 1.63 |
| Intersection influence: Traffic Signal [PRC] | 0.564 | 1.73 | 0.641 | 1.73 | 0.639 | 1.73 | 0.606 | 1.67 |
| Intersection influence: Traffic Signal [SC] | 0.625 | 1.45 | 0.763 | 1.58 | 0.735 | 1.52 | 0.689 | 1.27 |
| Darkness, lighted [PRC] | -0.713 | -1.91 | -0.774 | -1.83 | -0.819 | -1.92 | -0.611 | -1.37 |
| Darkness, lighted [SC] | -1.532 | -2.36 | -1.574 | -2.22 | -1.632 | -2.31 | -1.329 | -1.72 |
| Number of through lanes [SC] | 0.802 | 3.96 | 0.835 | 3.78 | 0.841 | 3.78 | 0.858 | 3.76 |
| Mist/light rain [SC] | 1.758 | 2.81 | 1.971 | 2.81 | 1.927 | 2.76 | 1.931 | 2.45 |

Notes: CV is coefficient of variation; K = 1, 2 refer to the volatility measures calculated using first and second 10-seconds driving data respectively; MC is minor crash; PRC is police-reportable crash; SC is most severe crash; (---) is Not Applicable.

**TABLE 4: Model Estimation Results for Crash Severity in Naturalistic Driving Environment (Second Specification) (Continued)**

| Variable Name | Fixed Parameter Multinomial Logit Model | | Random Parameter Multinomial Logit Model | | Random Parameter Multinomial Logit Model - Heterogeneity-in-Means | | Random Parameter Multinomial Logit Model - Heterogeneity-in-Means and Variances | |
|---|---|---|---|---|---|---|---|---|
| | β | t-stat | β | t-stat | β | t-stat | β | t-stat |
| **Fault Status** | | | | | | | | |
| Subject driver on fault [SC] | -1.114 | -2.86 | -1.258 | -2.95 | -1.391 | -3.18 | -1.448 | -3.27 |
| **Heterogeneity-in-means** | | | | | | | | |
| CV: longitudinal acceleration, K=1 [MC]: No secondary tasks | --- | --- | --- | --- | -1.799 | -2.76 | -1.791 | -2.83 |
| Mean Speed (km/h), K = 1 [MC]: Subject driver on fault | --- | --- | --- | --- | -0.024 | -1.77 | -0.028 | -2.11 |
| **Heterogeneity-in-variances** | | | | | | | | |
| Mean Speed (km/h), K = 1 [MC]: Darkness, lighted | --- | --- | --- | --- | --- | --- | 1.578 | 1.88 |

Notes: CV is coefficient of variation; K = 1, 2 refer to the volatility measures calculated using first and second 10-seconds driving data respectively; MC is minor crash; PRC is police-reportable crash; SC is most severe crash; (---) is Not Applicable.

**TABLE 5: Population Level Distribution Effects of the Random Parameters Logit,
Random Parameters Logit with Heterogeneity-in-Means, and Random Parameters Logit
with Heterogeneity-in-Means and Variances**

| Variables | Population-level Directional Heterogeneity | |
|---|---|---|
| | **Below 0** | **Above 0** |
| *Model Specification 1* | | |
| **Random Parameter Logit** | | |
| Mean Speed (km/h) [PRC] | 43.40% | 56.60% |
| CV: longitudinal deceleration [MC] | 23.12% | 76.88% |
| **Random Parameter Logit (Heterogeneity-in-Means)** | | |
| Mean Speed (km/h) [PRC] | 33.40% | 66.60% |
| CV: longitudinal deceleration [MC] | 0.95% | 99.05% |
| **Random Parameter Logit (Heterogeneity-in-Means & Variances)** | | |
| Mean Speed (km/h) [PRC] | 24.80% | 75.20% |
| CV: longitudinal deceleration [MC] | 2.48% | 97.52% |
| *Model Specification 2* | | |
| **Random Parameter Logit** | | |
| Mean Speed (km/h), K = 1 [MC] | 25.00% | 75% |
| CV: longitudinal acceleration, K=1 [MC] | 47.49% | 52.51% |
| **Random Parameter Logit (Heterogeneity-in-Means)** | | |
| Mean Speed (km/h) [PR] | 18.64% | 81.36% |
| CV: longitudinal deceleration [MC] | 34.64% | 65.36% |
| **Random Parameter Logit (Heterogeneity-in-Means & Variances)** | | |
| Mean Speed (km/h) [PRC] | 11.85% | 88.15% |
| CV: longitudinal deceleration [MC] | 32.76% | 67.24% |

Notes: CV is coefficient of variation; K = 1 refers to the volatility measures calculated using first 10-seconds driving data; MC is minor crash; PRC is police-reportable crash; SC is most severe crash.

**TABLE 6: Direct Marginal Effects of the Random Parameters Heterogeneity-in-Means and Variances Models**

| Variable Name | Specification 1 Random Parameter Multinomial Logit Model - Heterogeneity-in-Means and Variances | | | | Specification 2 Random Parameter Multinomial Logit Model - Heterogeneity-in-Means and Variances | | | |
|---|---|---|---|---|---|---|---|---|
| | **Mean** | **SD** | **Min** | **Max** | **Mean** | **SD** | **Min** | **Max** |
| **Volatility based on entire 30-seconds driving data** | | | | | | | | |
| CV: longitudinal acceleration [SC] | 0.0418 | 0.0031 | 0.0000 | 0.6217 | --- | --- | --- | --- |
| CV: longitudinal deceleration [SC] | 0.2654 | 0.0176 | 0.0000 | 3.6687 | --- | --- | --- | --- |
| CV: lateral acceleration [SC] | 0.0614 | 0.0041 | 0.0000 | 0.8858 | --- | --- | --- | --- |
| CV: lateral deceleration [SC] | 0.0237 | 0.0015 | 0.0000 | 0.2537 | --- | --- | --- | --- |
| Mean Speed (km/h) [SC] | 0.0788 | 0.0055 | 0.0000 | 1.0008 | --- | --- | --- | --- |
| CV: longitudinal acceleration [PRC] | 0.0345 | 0.0018 | 0.0002 | 0.5233 | --- | --- | --- | --- |
| CV: longitudinal deceleration [PRC] | 0.2286 | 0.0124 | 0.0002 | 2.9422 | --- | --- | --- | --- |
| CV: lateral acceleration [PRC] | 0.0488 | 0.0023 | 0.0007 | 0.6378 | --- | --- | --- | --- |
| Mean Speed (km/h) [PRC] | 0.0553 | 0.0041 | -0.2498 | 0.3770 | --- | --- | --- | --- |
| CV: longitudinal acceleration [MC] | 0.0701 | 0.0015 | 0.0021 | 0.4483 | --- | --- | --- | --- |
| CV: longitudinal deceleration [MC] | 0.1751 | 0.0046 | -0.0284 | 0.6858 | --- | --- | --- | --- |
| CV: lateral acceleration [MC] | 0.0894 | 0.0017 | 0.0069 | 0.3434 | --- | --- | --- | --- |
| Mean Speed (km/h) [MC] | 0.1249 | 0.0032 | 0.0002 | 0.5386 | --- | --- | --- | --- |
| **Volatility Based on Segmented Driving Data** | | | | | | | | |
| CV: longitudinal acceleration, K=1 [SC] | --- | --- | --- | --- | 0.0488 | 0.0031 | 0.0000 | 0.4011 |
| CV: longitudinal deceleration, K=1 [SC] | --- | --- | --- | --- | 0.0521 | 0.0034 | 0.0000 | 0.5015 |
| CV: lateral acceleration, K = 1 [SC] | --- | --- | --- | --- | 0.0578 | 0.0037 | 0.0000 | 0.5474 |
| CV: lateral deceleration, K = 1 [SC] | --- | --- | --- | --- | 0.0373 | 0.0023 | 0.0000 | 0.3832 |
| Mean Speed (km/h), K = 1 [SC] | --- | --- | --- | --- | 0.1055 | 0.0071 | 0.0000 | 1.0481 |
| CV: lateral deceleration, K = 2 [SC] | --- | --- | --- | --- | 0.1991 | 0.0135 | 0.0000 | 1.9227 |
| CV: longitudinal acceleration, K=1 [PR] | --- | --- | --- | --- | 0.0526 | 0.0018 | 0.0009 | 0.3077 |
| CV: longitudinal deceleration, K=1 [PR] | --- | --- | --- | --- | 0.1133 | 0.0046 | 0.0001 | 0.6570 |
| CV: longitudinal deceleration, K = 2 [PR] | --- | --- | --- | --- | 0.2175 | 0.0091 | 0.0004 | 1.3306 |
| CV: lateral acceleration, K = 2 [PR] | --- | --- | --- | --- | 0.0459 | 0.0016 | 0.0010 | 0.2553 |
| Mean Speed (km/h), K = 1 [PR] | --- | --- | --- | --- | 0.1137 | 0.0048 | 0.0004 | 0.7044 |
| CV: longitudinal acceleration, K=1 [MC] | --- | --- | --- | --- | 0.0745 | 0.0031 | -0.1564 | 0.3250 |
| CV: lateral deceleration, K = 1 [MC] | --- | --- | --- | --- | 0.0753 | 0.0018 | 0.0000 | 0.3188 |
| Mean Speed (km/h), K = 1 [MC] | --- | --- | --- | --- | 0.0718 | 0.0028 | -0.1105 | 0.4709 |
| **Secondary Tasks and Durations** | | | | | | | | |
| Cell phone, Texting [PR] | 0.0053 | 0.0011 | 0.0000 | 0.3689 | 0.0092 | 0.0019 | 0.0000 | 0.3530 |
| Duration in seconds of 1st secondary task [MC] | 0.0515 | 0.0026 | 0.0000 | 0.4973 | --- | --- | --- | --- |
| Duration in seconds of 2nd secondary task [MC] | 0.0189 | 0.0019 | 0.0000 | 0.4388 | --- | --- | --- | --- |
| **Driver Hand Status and Maneuver Judgement** | | | | | | | | |
| Both hands on wheel [MC] | -0.0449 | 0.0019 | -0.1489 | 0.0000 | -0.0448 | 0.0021 | -0.1784 | 0.0000 |
| No hands on wheel [SC] | 0.0037 | 0.0012 | 0.0000 | 0.3802 | 0.0068 | 0.0019 | 0.0000 | 0.5874 |
| Unsafe but legal maneuver [MC] | 0.0135 | 0.0016 | 0.0000 | 0.2072 | 0.0166 | 0.0020 | 0.0000 | 0.2469 |

Notes: CV is coefficient of variation; K = 1, 2 refer to the volatility measures calculated using first and second 10-seconds driving data respectively; MC is minor crash; PRC is police-reportable crash; SC is most severe crash; (---) is Not Applicable; SD is standard deviation.

**TABLE 6: Direct Marginal Effects of the Random Parameters Heterogeneity-in-Means and Variances Models (Continued)**

| Variable Name | Specification 1 Random Parameter Multinomial Logit Model - Heterogeneity-in-Means and Variances | | | | Specification 2 Random Parameter Multinomial Logit Model - Heterogeneity-in-Means and Variances | | | |
|---|---|---|---|---|---|---|---|---|
| | Mean | SD | Min | Max | Mean | SD | Min | Max |
| **Locality and Environmental Factors** | | | | | | | | |
| Open residential area [MC] | 0.0089 | 0.0016 | 0.0000 | 0.3446 | 0.0119 | 0.0021 | 0.0000 | 0.4331 |
| School zone [SC] | 0.0033 | 0.0007 | 0.0000 | 0.1651 | 0.0065 | 0.0013 | 0.0000 | 0.2706 |
| Intersection influence: Traffic Signal [PRC] | 0.0087 | 0.0009 | 0.0000 | 0.1875 | 0.0128 | 0.0012 | 0.0000 | 0.1490 |
| Intersection influence: Traffic Signal [SC] | 0.0094 | 0.0011 | 0.0000 | 0.1650 | 0.0099 | 0.0012 | 0.0000 | 0.1649 |
| Darkness, lighted [PRC] | -0.0076 | 0.0009 | -0.1979 | 0.0000 | -0.0083 | 0.0008 | -0.1349 | 0.0000 |
| Darkness, lighted [SC] | -0.0087 | 0.0016 | -0.3545 | 0.0000 | -0.0068 | 0.0011 | -0.2588 | 0.0000 |
| Number of through lanes [SC] | 0.0529 | 0.0037 | 0.0000 | 0.7113 | 0.0691 | 0.0045 | 0.0000 | 0.8127 |
| Mist/light rain [SC] | 0.0068 | 0.0015 | 0.0000 | 0.3834 | 0.0092 | 0.0020 | 0.0000 | 0.4605 |
| **Fault Status** | | | | | | | | |
| Subject driver on fault [SC] | -0.0882 | 0.0066 | -0.7914 | 0.0000 | -0.0461 | 0.0031 | -0.3462 | 0.0000 |

Notes: MC is minor crash; PRC is police-reportable crash; SC is most severe crash; SD is standard deviation.